\documentclass[aps,twocolumn,superscriptaddress]{revtex4-2}
\usepackage{amsmath,bm,graphicx,amssymb,epsfig,dsfont,xcolor}

\begin{document}

\title{Crystallization of the transdimensional electron liquid}

\author{Igor V. Bondarev}
\affiliation{Department of Mathematics \& Physics, North Carolina Central University, Durham, NC 27707, USA}

\author{Alexandra Boltasseva}
\affiliation{Elmore Family School of Electrical \& Computer Engineering, Purdue Quantum Science \& Engineering Institute, and Birck Nanotechnology Center, West Lafayette, IN 47907, USA}
\affiliation{School of Materials Engineering, Purdue University, West Lafayette, IN 47907, USA}

\author{Jacob B. Khurgin}
\affiliation{Department of Electrical \& Computer Engineering, Whiting School of Engineering, Johns Hopkins University, Baltimore, MD 21218, USA}

\author{Vladimir M. Shalaev}
\affiliation{Elmore Family School of Electrical \& Computer Engineering, Purdue Quantum Science \& Engineering Institute, and Birck Nanotechnology Center, West Lafayette, IN 47907, USA}

\begin{abstract}
Wigner crystallization of free electrons at room temperature is explored for a new class of metallic ultrathin (transdimensional) materials whose properties can be controlled by their thickness. Our calculations of the critical electron density, temperature and the melting curve show that by reducing the material thickness one can Wigner-crystallize free electrons at room temperature to get them pinned onto a two-dimensional triangular lattice of a supersolid inside of the crystalline material. Such a solid melts and freezes reversibly with increase and decrease of electron doping or temperature, whereby its resistivity behaves opposite to the free electron gas model predictions.
\end{abstract}

\maketitle


Condensed matter systems with strong electron correlations have long been in the focus of theoretical and experimental studies due to their unique physical properties~\cite{Wigner,Anderson,Kondo,Kleiner,Williams,Kosterlitz,PF1974,LozovikYudson,Grimes,FukuyamaPA,LozovikFA,Shayegan}. These studies have now evolved into a vibrant field of quantum nanomaterials to explore correlated electron systems of reduced dimensionality for remarkable phenomena such as high-temperature ($T$) quantum phase transitions~\cite{Fogler14,Suris16}, superconductivity~\cite{LozovikUFN18,Snoke21,Snoke24}, unconventional magnetism~\cite{Zhou20}, and a variety of metal-insulator transitions (MITs) including quantum- and disorder-related Anderson localization~\cite{Segev,Pruneri}, Kondo effect~\cite{Kouwenhoven,Otsuka}, Wigner crystallization~\cite{Bockrath,Ashoori,Ilani,FengWang2020,FengWang2021,Smolenski,Park,FengWang2024a,FengWang2024b,Yazdani2024} and beyond~\cite{MakShan,ShanMak,Cui,MacDonald,Sushkov}. These effects are studied in various strongly correlated materials for electrons, excitons and their complexes~\cite{MacDonald,Balatsky,Bondarev2020,Bondarev2021,Lu2022,Dai}, particularly in the low-dimensional regime, in systems such as semiconductor quantum wells, graphene and transition metal dichalcogenides (TMDCs).

One of the most interesting MIT phenomena is the electron Wigner crystal formation~\cite{Wigner}---the longest anticipated exotic correlated phase of metals and metallic compounds that has intrigued physicists since 1934~\cite{Monarkha}. In this phase free electrons crystallize in metals on a periodic lattice to form a solid made of a superlattice of electrons inside of a crystalline material. Electrons become pinned (frozen) periodically when their potential repulsion energy exceeds both the mean kinetic energy per particle and the energy of thermal fluctuations, with their density and $T$ not to exceed dimension-dependent critical values~\cite{PF1974}. Such a solid melts and freezes up reversibly with increase and decrease of $T$, respectively, and its resistivity $T$-dependence is opposite to the free electron gas model predictions. In spite of a large body of research, achieving and observing electron Wigner crystallization remains an outstanding challenge that requires high quality, structurally stable metallic compounds with tailorable electronic response and low disorder to succeed. Thus far, signatures of electronic Wigner crystallization were observed indirectly in 2D electron gas systems under high magnetic fields~\cite{Shayegan,Ashoori} and in twisted bilayer TMDC moir\'{e} superlattices (generalized Wigner crystals~\cite{FengWang2020,FengWang2021,FengWang2024a,FengWang2024b,ShanMak,MakShan,Cui}). Only recently, the first microscopic images to prove charge excitations in one-dimensional (1D)~\cite{Ilani,Bockrath}, non-zero magnetic field 2D~\cite{Yazdani2024} and generalized 2D Wigner crystals were reported~\cite{FengWang2021,FengWang2024a,FengWang2024b}. The last two are different from the Wigner's electron crystal concept as the "crystallization" there is due either to magnetic localization or to moir\'{e} potential trapping of electrons instead of their Coulomb repulsion. To date, the zero-field electron crystallization has been observed indirectly in semiconducting TMDC monolayers~\cite{Smolenski} and in untwisted homobilayers~\cite{Park} by monitoring exciton photoluminescence intensity. Specifically, an extra peak was detected that could originate from the exciton Umklapp scattering by the 2D electron lattice formed below the Wigner crystal melting point ($\sim\!10$~K)~\cite{Smolenski}. All these studies require low $T$ and external means to reduce electron mobility (magnetic field, moir\'{e} potential). Thus, the observation of the Wigner's prediction in conventional materials remains elusive.

With current nanofabrication technology development, an exciting opportunity to study strongly correlated phenomena is offered by the so-called transdimensional (TD) material platform~\cite{MRStransdim}. Originally proposed in the field of nanoplasmonics~\cite{BoltShal2019,GDA,ManGDA,Atwater}, these ultra-thin---between 2D and 3D---materials are expected to support strong electron correlations and could potentially enable quantum phenomena such as Wigner crystallization~\cite{BoltBondShal2025}. Metallic and semimetallic TD compounds can have thicknesses of only a few atomic layers and show unprecedented tailorability of their electromagnetic (EM) response~\cite{BondShal2017,BondMousShal2018,BondMousShal20,Bond2023,BiehsBond24}. This includes unusually strong dependence on structural parameters such as thickness (number of atomic monolayers), composition (stoichiometry, doping), strain and surface termination compared to conventional thin films, as well as extreme sensitivity to external optical and electrical stimuli. Recently, epitaxial TD films of transition metal nitrides (TMNs) such as TiN, ZrN and HfN, have been studied extensively~\cite{Naik} and demonstrated their confinement-induced nonlocal EM response as well as new associated physical effects~\cite{Shah2022,BiehsBond23,Das2024}. However, while quite a few confinement-induced plasmonic effects have been reported for TD films experimentally~\cite{Shah2022,Salihoglu2023,Das2024}, until very recently TD materials have not been used to explore strongly correlated electron regimes. The first experimental evidence for plasmonic behavior breakdown and related MIT was reported recently for room-$T$ HfN films decreasing in thickness $d$ to become a transparent dielectric at $d\!=\!2$~nm~\cite{Das2024}. The unique possibility to observe the reversible MIT due to electron Wigner crystallization in vertically confined planar metallic structures not only provides insights into strong electron correlation phenomena but is also attractive for nanophotonics applications. When free electrons crystallize into a superlattice, the TD film turns into an optically transparent dielectric. When the electron solid melts, the film restores its plasmonic response. The exploration of Wigner crystallization in TD materials opens a new avenue for the realization of optical modulation and switching with this new photonic material platform.

\begin{figure}[t]
\includegraphics[width=0.48\textwidth]{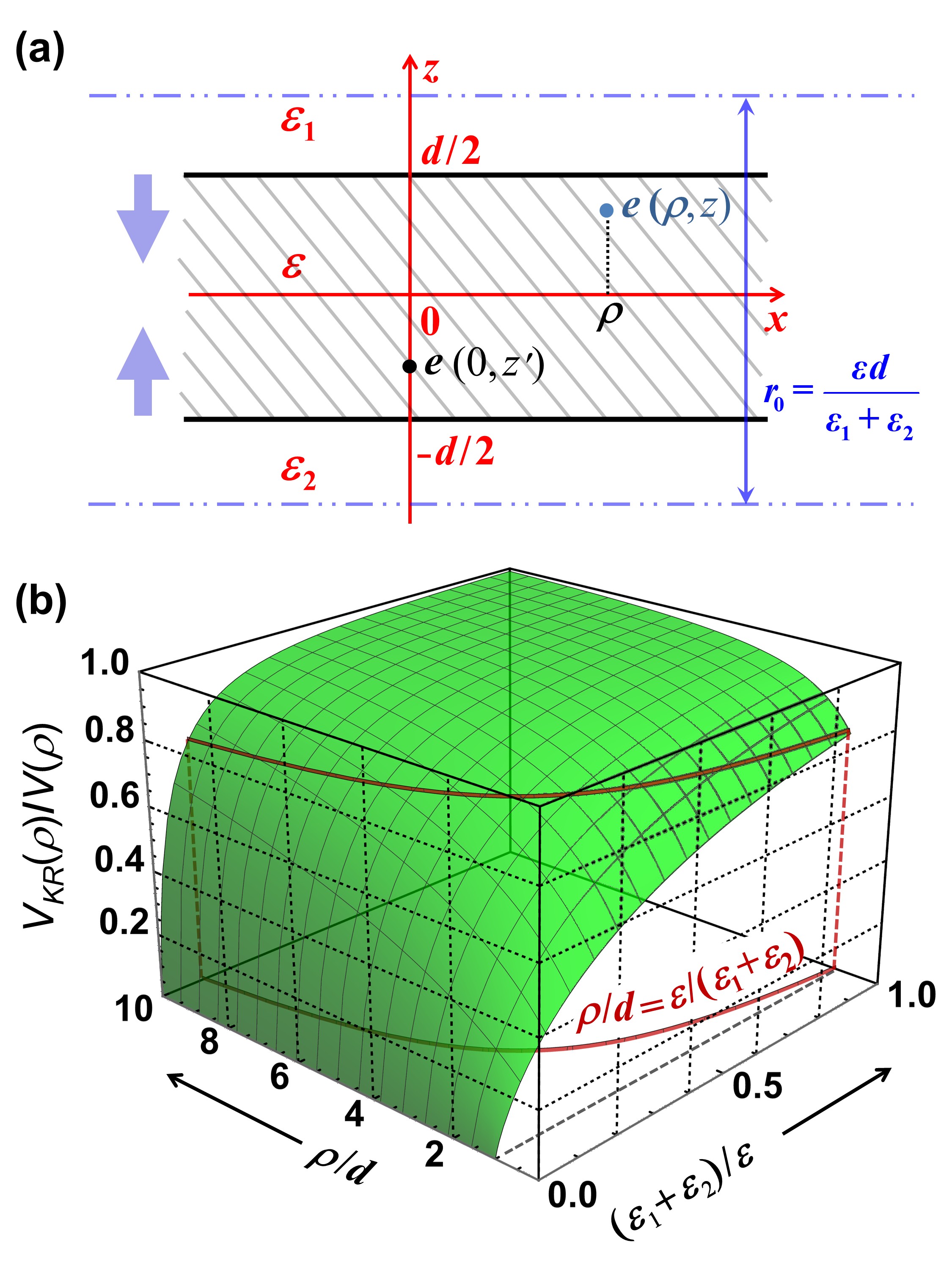}\vspace{-0.5cm}
\caption{(a)~Schematic to show the geometry of the KR electrostatic potential (\ref{VKR}) for a pair of electrons confined in the ultrathin TD slab of decreasing thickness. (b)~KR interaction potential (\ref{VKR}) normalized (divided) by the 2D Coulomb interaction potential given by Eq.~(\ref{VKR}) in the limit $d\!\rightarrow\!0$.}
\label{fig1}\vspace{-0.25cm}
\end{figure}

Here, we develop a theory to generalize the Platzman-Fukyuama (PF) model of the Wigner crystal formation in free-standing 2D electron gas systems~\cite{PF1974}, to the practical case of TD plasmonic materials. Our calculations of the critical density, temperature, and the actual melting surface to identify the Wigner crystal phase in the available (broader) parameter space, show that TD materials offer a unique possibility. Specifically, we show that it is possible to crystallize electrons even at room $T$ by simply reducing the thickness of the material. By reducing the thickness one decreases the electron density and, at the same time, enhances the inter-electron repulsive potential due to the vertical confinement. By doing this in a controllable manner it is possible to make the repulsive potential exceed the mean single-electron kinetic energy so that a stable room-$T$ Wigner solid can be formed.


As first formulated by Platzman and Fukuyama~\cite{PF1974}, an ensemble of repulsively interacting particles (or quasiparticles) is expected to form a Wigner crystal lattice when its average pair potential interaction energy $\langle V\rangle$ exceeds the average kinetic energy per particle $\langle K\rangle$. Then the ratio $\Gamma_0\!=\!\langle V\rangle/\langle K\rangle\!>\!1$ (referred to as the PF ratio below) represents the phase diagram (melting curve) of the process if $\Gamma_0$ is known. The PF model describes an idealized 2D electron gas system, free-standing in air, where $\langle V\rangle\!=\!e^2/\langle\rho\rangle\!=\!e^2\sqrt{\pi n}$ with $n$ being the 2D electron density defined by mean in-plane inter-electron distance $\langle\rho\rangle$ through the constraint $\pi\langle\rho\rangle^2\!=\!1/n$. Since liquids do not support transverse vibrational modes and solids do, the model uses an assumption that the transverse vibrational mode instability of the electron Wigner crystal signals the onset of its melting. This leads to $\Gamma_0\!\approx\!3$ at melting~\cite{PF1974}, with zero-$T$ critical density scaling unit $n_c\!=\!1/(\pi a_B^2\Gamma_0^2)\!\approx\!5\!\times\!10^{15}\,$cm$^{-2}$ if calculated using the 2D Bohr radius $a_B\!=\!\hbar^2\!/(2me^2)\!=\!0.529/2\;\mbox{\AA}$. The critical temperature scaling unit given under the classical energy equipartition by $\Gamma_0\!=\!e^2\sqrt{\pi n_c}/(k_BT_c)$, equals $k_BT_c\!=\!2Ry/\Gamma_0^2\!\approx\!6$~eV with the 2D Rydberg constant $Ry\!=\!e^2/(2a_B)\!=\!27.25$~eV.

For metallic TD films of thickness $d$, the PF ratio is
\begin{equation}
\Gamma=\frac{\langle V_{K\!R}\rangle}{\langle K\rangle}.
\label{PFratio}
\end{equation}
Here
\begin{equation}
V_{K\!R}(\rho)=\frac{e^2\pi}{(\varepsilon_1+\varepsilon_2)r_0}\left[H_{0}\Big(\frac{\rho}{r_0}\Big)-N_{0}\Big(\frac{\rho}{r_0}\Big)\right]
\label{VKR}
\end{equation}
is the repulsive Keldysh-Rytova (KR) interaction potential~\cite{KeldyshRytova} written in Gaussian units as a difference of the $0$-order Struve ($H_0$) and Neumann ($N_0$, \emph{aka} Bessel $Y_0$~\cite{Rubio11}) special functions, where $r_0\!=\!\varepsilon d/(\varepsilon_1+\varepsilon_2)\!=\!2\pi\alpha_{2D}$ is the screening length with $\alpha_{2D}$ representing the in-plane polarizability of 2D material~\cite{Rubio11}. This is the electrostatic repulsive interaction energy of a pair of electrons separated by the in-plane distance $\rho$ and confined vertically in the interior of the optically dense TD film with the positive background permittivity $\varepsilon\!>\!\varepsilon_1,\varepsilon_2$ of superstrate and substrate as shown in Fig.~\ref{fig1}~(a). The KR potential indicates that in such optically dense ultrathin planar systems the vertical electron confinement leads to the effective dimensionality reduction from 3D to 2D, with the $z$-coordinate of the potential replaced by new parameter $d$ representing the vertical size. The potential $V_{K\!R}$ can be shown to go logarithmically with $\rho$ for $d\!\ll\!\rho\!\ll\!r_0$ and fall off as $1/\rho$ for $\rho\!\gg\!r_0$~\cite{KeldyshRytova,Rubio11}; see Fig.~\ref{fig1}~(b). It can be accurately approximated by elementary functions as
\begin{eqnarray}
V_{K\!R}(\rho)\approx\frac{2e^2}{(\varepsilon_1\!+\varepsilon_2)r_0}\!\left[\ln\!\Big(1\!+\!\frac{r_0}{\rho}\Big)\!+(\ln2\!-\!\gamma)e^{-\rho/r_0}\!\right]~~
\label{VKRelemfunc}
\end{eqnarray}
($\gamma\!\approx\!0.577$ is the Euler-Mascheroni constant). This expression was originally proposed for monolayer semiconductors~\cite{Rubio11}. It can be seen from Eq.~(\ref{VKRelemfunc}) that the PF model is inappropriate for the description of the finite-thickness TD films as the standard 2D Coulomb coupling is not the case there. In the pure 2D regime ($d\!\rightarrow\!0$) where it is set to work, the PF ratio is still to be multiplied by $2/(\varepsilon_1\!+\varepsilon_2)$ to include the substrate and superstrate for realistic atomically thin but optically dense materials.

\begin{figure}[t]
\hskip-0.35cm\includegraphics[width=0.48\textwidth]{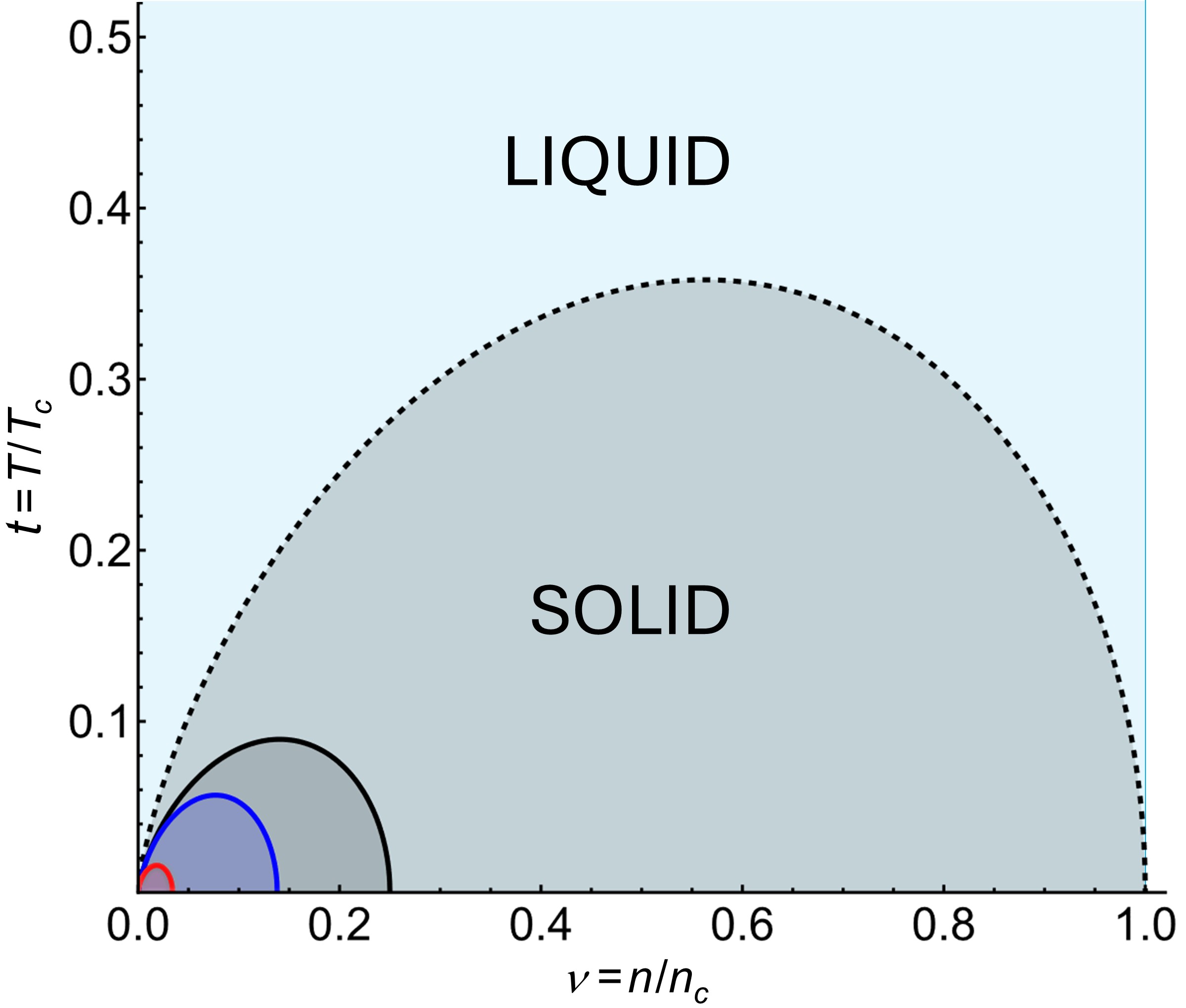}\vspace{-0.25cm}
\caption{Solid-liquid phase diagrams for TD films as compared to 2D free-standing PF model (black dashed line). Red, blue, and black lines enclose Wigner solid phases (shaded accordingly) for air($\varepsilon_1\!=\!1$)/TiN($\varepsilon\!=\!9$)/MgO($\varepsilon_2\!=\!3$) TD systems of $d\!=\!1\,\mbox{nm},\,0.1\,$nm, and $d\!\rightarrow\!0$, respectively, as given by Eq.~(\ref{MeltingSurface}).}\vspace{-0.25cm}
\label{fig2}
\end{figure}

The mean electron kinetic energy per particle can be calculated analytically for all $T\!<\!T_F$, the Fermi temperature of metals ($\sim\!10^5\,$K) by integrating over the 2D reciprocal space~\cite{Supplementary}, to yield
\begin{equation}
\langle K\rangle=\frac{\pi n\hbar^2}{2m}\!\left[1\!+\!\frac{1}{3}\Big(\frac{m k_BT}{n\hbar^2}\Big)^{\!2}\right]\!=k_BT_{c\,}\nu\Big(1+\frac{\pi^2t^2}{12\nu^2}\Big).
\label{K}
\end{equation}
Here, $\nu\!=\!n/n_c$ and $t\!=\!T/T_c$ are the electron density and temperature, respectively, made dimensionless using the critical density and critical temperature scaling units above for convenience of comparison with the PF model.

Plugging Eqs.~(\ref{VKR}) and (\ref{K}) in Eq.~(\ref{PFratio}) leads to
\begin{equation}
\frac{\Gamma}{\Gamma_0}=\frac{F(d,\varepsilon,\varepsilon_{1,2})}{\nu\big[1+\pi^2t^2/(12\nu^2)\big]},
\label{PFratiogen}
\end{equation}
where
\begin{equation}
F(d,\varepsilon,\varepsilon_{1,2})\!=\!\pi\Gamma_0\frac{H_{0\!}\big[\Gamma_0/(\bar{r}_0\sqrt{\nu})\big]\!-\!N_{0\!}\big[\Gamma_0/(\bar{r}_0\sqrt{\nu})\big]}{(\varepsilon_1+\varepsilon_2)\,\bar{r}_0}
\label{F}
\end{equation}
is the dimensionless function of the TD film parameters with $1/\bar{r}_0\!=\!(\varepsilon_1\!+\varepsilon_2)a_B/(\varepsilon d)\!=\!a_B/(2\pi\alpha_{2D})$. For $d\!\rightarrow\!0$, the $F$ function power series expansion at infinity does not contain even-degree terms, which makes the first-order series expansion term quite a good approximation when $d$ is small enough. Then Eq.~(\ref{F}) results in $F\!=\!2\sqrt{\nu}/(\varepsilon_1\!+\varepsilon_2)$ for any $\varepsilon\!>\!\varepsilon_{1,2}$, and Eq.~(\ref{PFratiogen}) subject to $\Gamma/\Gamma_0\!=\!1$ yields the constraint $t\!=\!2\sqrt{3\nu\,[2\sqrt{\nu}/(\varepsilon_1\!+\!\varepsilon_2)\!-\nu]}/\pi$ in the $(\nu,t)$ two-coordinate space. This is the 'melting curve' to divide the $(\nu,t)$ plane into the regions of the solid phase formed by the electron superlattice and conventional liquid phase of the free electron system. A simple extreme value analysis reveals the only point of maximum for this curve, $\nu_0\!=\!2.25/(\varepsilon_1\!+\varepsilon_2)^2$ and $t_0\!=\!4.5\pi^{-1}\!/(\varepsilon_1\!+\varepsilon_2)^2$, in the square-root domain $0\!\le\!\nu\!\le\!4/(\varepsilon_1\!+\varepsilon_2)^2$ which with the $0\!\le\!t\!<\!t_0$ condition encloses the electron Wigner crystal phase. For example, for air$\,(\varepsilon_1\!=\!1)$/MgO$\,(\varepsilon_2\!=\!3)$ super\-strate/substrate atomically thin ($d\!\rightarrow\!0$) TD systems, $n_0\!\approx\!7\!\times\!10^{14}\,$cm$^{-2}$, $T_0\!\approx\!6110$~K and $n\!\le\!1.25\!\times\!10^{15}\,$cm$^{-2}$. For $\varepsilon_1\!=\!\varepsilon_2\!=\!1$ the curve turns into that of PF model to enclose the region $0\!\le\!\nu\!\le\!1$, $0\!\le\!t\!\lesssim\!0.4$ of the Wigner crystal phase for 2D electron system free-standing in air~\cite{PF1974}.

In the most general case of the ultrathin TD films of finite-thickness, Eq.~(\ref{PFratiogen}) yields the melting surface in the $(d,\nu,t)$ three-coordinate space
\begin{equation}
t=\!\frac{2}{\pi}\sqrt{3\nu\big[F(d,\varepsilon,\varepsilon_{1,2})-\nu\big]}\,.
\label{MeltingSurface}
\end{equation}
This turns into the PF melting curve when projected on the $d\!=\!0$ plane with $\varepsilon_1\!=\varepsilon_2\!=\!1$. In the opposite limit, raising $d$ makes the square-root argument negative~\cite{Supplementary}, and the Wigner crystal phase is rendered impossible.

\begin{figure}[t]
\hskip-0.35cm\includegraphics[width=0.48\textwidth]{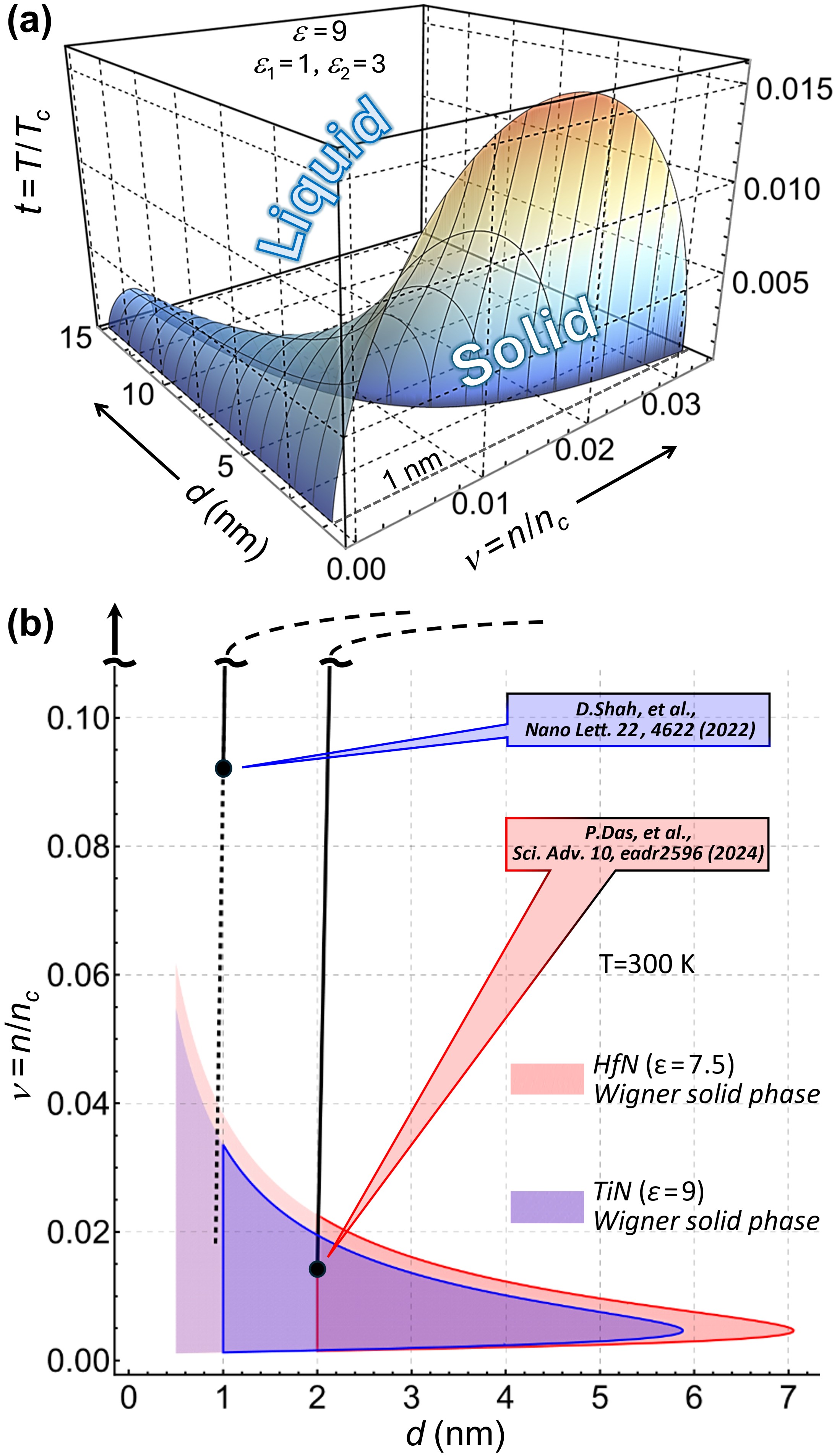}\vspace{-0.35cm}
\caption{(a)~Melting surface given by Eq.~(\ref{MeltingSurface}) for air/TiN/MgO TD film. (b)~Horizontal crosscut of (a) to show Wigner crystal phase at room $T$ (light blue) and same for air/HfN/MgO (light red). Solid lines confine the areas accessed experimentally. Black dots show the lowest electron densities reported ($7.4\!\times\!10^{13}$cm$^{-2}$ and $4.5\!\times\!10^{14}$cm$^{-2}$ for $2\,$nm thick HfN and $1\,$nm thick TiN; thickness dependences traced by black lines).}\vspace{-0.45cm}
\label{fig3}
\end{figure}

The features above-described for the TD film melting surfaces can be seen in Figs.~\ref{fig2} and~\ref{fig3}, obtained numerically from Eq.~(\ref{MeltingSurface}). Figure~\ref{fig2} shows the curves calculated for fixed $d$ with $\varepsilon_1\!=\!1$, $\varepsilon\!=\!9$ and $\varepsilon_2\!=\!3$ (see Fig.~\ref{fig1}) corresponding to the air/TiN/MgO TD system~\cite{Shah2022}. The free-standing ($\varepsilon_1\!=\!\varepsilon_2\!=\!1$) zero-$d$ PF model curve is shown as well. The Wigner solid phases are bounded from the top by their respective melting curves and shaded accordingly. They can be seen to contract significantly not only with increasing $d$ but also for atomically thin films deposited on a dielectric substrate compared to the free standing case. Figure~\ref{fig3}~(a) shows the melting surface for the same TD system in the $(d,\nu,t)$ three-coordinate space. The solid phase expands drastically as $d\!\rightarrow\!0$ while still taking just a hundredth of the parameter space of the free-standing PF model case (cf. Fig.~\ref{fig2}).

The range of parameters for the room-$T$ Wigner crystal phase in the TD plasmonic film systems can be seen in Fig.~\ref{fig3}~(b). Shown in light blue and red are the horizontal $T\!=\!300\,$K crosscuts of the air/TiN/MgO and air/HfN/MgO ($\varepsilon\!=\!7.5$) melting surfaces. Blue and red solid lines confine the parameter ranges previously accessed experimentally~\cite{Das2024,Shah2022}. Black lines trace the $n(d)$ dependences reported (in units of $n_c$), with black dots to indicate the lowest $n$ measured: $7.4\!\times\!10^{13}\,$cm$^{-2}$ for $2\,$nm thick HfN film and $4.5\!\times\!10^{14}\,$cm$^{-2}$ for $1\,$nm thick TiN film (estimated from plasma frequency measured). As can be seen, free electrons of $2\,$nm thick HfN film are expected to be in the stable Wigner solid state as the electron lattice correlation energy of $k_BT_c\!\approx\!67\,$meV, given by Eq.~(\ref{MeltingSurface}) for this case, exceeds greatly typical room-$T$ thermal fluctuation energies in 2D materials ($\lesssim\!10\,$meV~\cite{Balandin}). Thus, the plasmonic breakdown and MIT are unavoidable just as observed~\cite{Das2024}. A tiny increase of $d$ causes $n$ (or~$\nu$) to rapidly increase ($n_{2D}\!=\!n_{3D}d$), in which case the HfN system leaves the solid phase region as Fig.~\ref{fig3}~(b) shows and melts out to restore its plasmonic properties. Clearly, the effect is reversable. Figure~\ref{fig3}~(b) also shows that due to higher $n$ the TiN plasmonic breakdown effect cannot occur for $d\!\ge\!1\,$nm, in agreement with experiment~\cite{Shah2022}. However, removing just a monolayer of material could shift the TiN electron system to the appropriate region of parameter space for crystallization (black dotted line). Once it is there, not only the increase of $T$ but also the increase of $n$ (due to the electrostatic doping) can melt such a Wigner solid reversibly as can be seen from Fig.~\ref{fig3}~(a).

Signatures of electron Wigner crystallization in semiconductor TMDC monolayers were recently observed indirectly in zero magnetic field by monitoring an extra exciton photoluminescence resonance interpreted as being due to the exciton Umklapp scattering by the 2D electron lattice formed at $T\!\lesssim\!10\,$~K~\cite{Smolenski}. The effect was reported for $n\!\sim\!10^{11}\,$cm$^{-2}$, or $\nu\!=\!n/n_c\!\approx\!2\!\times\!10^{-5}$ in terms of our theory. For such small $\nu$, the $d$-dependence in Eq.~(\ref{F}) cancels out completely and only the substrate-superstrate dielectric factor remains for the 2D interface of \emph{h}-BN material the TMDC monolayer is embedded in. With $\varepsilon_1\!=\!\varepsilon_2\!\approx\!5.87$ for bulk \emph{h}-BN~\cite{Snoke21,h-BN}, Eq.~(\ref{MeltingSurface}) takes the form $t\!=\!T/T_c\!\approx\!2\sqrt{6\nu^{3/2}/(\epsilon_1\!+\!\epsilon_2)}/\pi\!\approx\!0.5\,\nu^{3/4}$ to locate monolayer TMDCs at the very bottom left corner on the $\nu$-axis in Fig.~\ref{fig3}~(a) where the Wigner crystal phase is bounded by $T\!\lesssim\!10\,$K, or at the very bottom below the shaded areas in Fig.~\ref{fig3}~(b) where no room-$T$ crystal phase exists. Earlier zero-field 2D $p$-doped GaAs/AlGaAs experiments ($\varepsilon_1\!=\!\varepsilon_2\!\approx\!12.5$) fall into that region as well due to even lower $\sim\!10^{10}\,$cm$^{-2}$ carrier densities~\cite{Shayegan99}, yielding $T\!\lesssim\!1\,$K for the upper bound of the crystal phase, just as was observed experimentally. In sharp contrast to zero-$T$ theory predictions~\cite{Tanatar89}, at very low densities electrostatic repulsion tends to zero while kinetic energy per particle remains finite due to quantum fluctuations, whereby the potential-to-kinetic energy ratio $\Gamma$ drops down necessitating lower $T$ for crystallization (see Ref.~\cite{Supplementary} for details).

To conclude, TD materials offer a new approach to explore strong electronic correlations in quantum systems. The screening in TD metals and semimetals is greatly reduced as compared to their bulk counterparts. Metal-dielectric interface barriers are high enough ($\sim\!3\;$eV) for electron spill-out distances not to exceed just a few fractions of angstrom~\cite{Hirabayashi}. Contrary to artificial 2D superlattices including moir\'{e} systems, where in-plane transport is suppressed due to an effective mass (which can be quite large) but even tiny imperfections can lead to irreversible disorder-related Anderson localization, TD materials such as TMNs are less imperfection sensitive and so are more suitable for Wigner crystal formation~\cite{Das2024}. As a test, an in-plane static magnetic field can be used to reduce the number of electron translational degrees of freedom from two (in-plane motion) to one (in-plane motion along the magnetic field) and thus to change the Wigner crystallization picture while leaving the Anderson localization process intact. A variety of TMNs (TiN, ZrN, HfN, etc.), their ability to grow as high-quality ultrathin epitaxial films with controlled interfacial strain~\cite{Chen2023}, and their electron density sensitivity to material parameters, provide a rich playground for the realization of strongly correlated electron systems in different regimes~\cite{BoltBondShal2025}. Exploring the Wigner crystal feasibility with TD materials at room $T$ in zero magnetic field represents an entirely new direction in the research area of strongly correlated electron systems. It is expected to bring critical fundamental insights into the physics of strongly correlated phenomena to enable a new generation of tunable, reconfigurable and multifunctional devices for nanophotonics, optoelectronics, and advanced quantum technologies.

\begin{acknowledgments}
I.V.B. gratefully acknowledges support from the U.S. Army Research Office under award W911NF2310206. A.B. and V.M.S. acknowledge support from the U.S. Department of Energy, Office of Basic Energy Sciences, Division of Materials Sciences and Engineering under award DE-SC0017717.
\end{acknowledgments}

\end{document}


\title{\texttt{\huge{Supplementary Information}}\\[3cm]Crystallization of the Transdimensional Electron Liquid\\[0.5cm]~}

\author{Igor V. Bondarev}
\affiliation{Department of Mathematics \& Physics, North Carolina Central University, Durham, NC 27707, USA}

\author{Alexandra Boltasseva}
\affiliation{Elmore Family School of Electrical \& Computer Engineering, Purdue Quantum Science \& Engineering Institute, and Birck Nanotechnology Center, West Lafayette, IN 47907, USA}
\affiliation{School of Materials Engineering, Purdue University, West Lafayette, IN 47907, USA}

\author{Jacob B. Khurgin}
\affiliation{Department of Electrical \& Computer Engineering, Whiting School of Engineering, Johns Hopkins University, Baltimore, MD 21218, USA}

\author{Vladimir M. Shalaev}
\affiliation{School of Materials Engineering, Purdue University, West Lafayette, IN 47907, USA}

\vskip1.5cm

\begin{abstract}
\vskip1.5cm
Here we provide the technicalities of the analytical calculations for the mean number of particles and mean kinetic energy per particle for free electrons in transdimensional (TD) material systems. Detailed discussion supported with illustrations is also provided to include the domain of the melting surface equation in the $(d,\nu,t)$ three-coordinate space and the generalized PF ratio temperature dependence of the TD systems.
\end{abstract}

\maketitle

\vskip2.5cm

\tableofcontents

\newpage

\section{Introduction}

For optically dense planar nanostructures in the transdimensional (TD) regime, the electrostatic interaction potential of charge carriers confined is stronger than that in a homogeneous medium with the same dielectric permittivity due to the increased field contribution from outside dielectric environment with lower dielectric constant~\cite{Rubio11,Louie09}. This interaction is associated with the Keldysh-Rytova (KR) electrostatic interaction potential~\cite{Keldysh1979}, in which the thickness $d$ represents the size of the vertical electron confinement region in optically dense ultrathin planar systems, while the vertical coordinate ($z$-coordinate) dependence is gone. This corresponds to the effective dimensionality reduction from 3D to 2D, both in the coordinate space and in the momentum (reciprocal) space. Therefore, all sums and integrals over the momentum space presented below are 2D, and are calculated using the standard statistical physics ansatz~\cite{LandauStatPhys}
\begin{equation}
\sum_{\textbf{k}\,\in\,1st\,B.Z.}\!\!\!\!\!\!\!\cdots=\frac{S}{(2\pi)^2}\!\int\!d\textbf{k}\cdots=\frac{S}{(2\pi)^2}2\pi\!\!\int\!dkk\cdots
=\frac{S}{2\pi}\!\int_0^{\infty}\!\!\!\!\!d\epsilon\frac{k}{d\epsilon/dk}\cdots,
\label{ansatz}
\end{equation}
where summation is over the first Brillouin zone of an in-plane isotropic 2D electron system of surface area $S$.

\section{Mean kinetic energy per particle at low and moderate temperatures}

This is the quantum degeneracy regime of the electron gas. Due to high Fermi temperatures $T_F\;(\sim\!10^5\;\mbox{K})$ of typical metals, this regime spreads from the absolute zero upwards to exceed the room temperature of $\sim\!300\;$K by two to three orders of magnitude. With Eq.~(\ref{ansatz}), the mean kinetic energy per particle takes the following form
\begin{equation}
\langle K\rangle=\frac{2}{\langle N\rangle}\sum_{\textbf{k}\,\in\,1st\,B.Z.}\epsilon(\textbf{k})n_F(\textbf{k})=
\frac{1}{\pi n}\!\int_0^{\infty}\!\!\!\!\!d\epsilon\frac{k\epsilon}{d\epsilon/dk}n_F(\epsilon)\,.
\label{meanK}
\end{equation}
Here,
\begin{equation}
\langle N\rangle\!=\!2\sum_{\textbf{k}}n_F(\textbf{k})
\label{meanN}
\end{equation}
is the mean number of particles (electrons) in the system (factor of 2 is to account for the electron spin degeneracy),
\begin{equation}
n_F(\textbf{k})=\frac{1}{e^{\beta[\epsilon(\textbf{k})-\mu]}+1},\;\;\;\beta=\frac{1}{k_BT}
\label{nF}
\end{equation}
is the Fermi-particle distribution function with chemical potential $\mu$ to represent the many-particle system of quasi-free electrons with in-plane quasimomentum $\textbf{k}$, effective mass $m$, and kinetic energy
\begin{equation}
\epsilon(\textbf{k})=\frac{\hbar^2k^2}{2m},
\label{ek}
\end{equation}
and
\begin{equation}
n=\frac{\langle N\rangle}{S}
\label{n}
\end{equation}
is the electron surface density of the system. Note that the quasi-free electron (\emph{aka} ideal electron gas) approximation is totally legitimate for degenerate electron gas systems and works the better the greater the system density is~\cite{LandauStatPhys}.

Using the ansatz (\ref{ansatz}) together with Eqs.~(\ref{nF}) and (\ref{ek}) in Eq.~(\ref{meanN}) gives
\begin{equation}
\langle N\rangle=2\frac{S}{(2\pi)^2}2\pi\frac{m}{\hbar^2}\int_{0}^{\infty}\!\!\!\frac{d\epsilon}{e^{\beta(\epsilon-\mu)}+1},
\label{N1}
\end{equation}
which after the substitution of variables
\begin{equation}
z=\beta(\epsilon-\mu)
\label{substitution}
\end{equation}
becomes
\begin{equation}
\langle N\rangle=2\frac{S}{(2\pi)^2}2\pi\frac{m}{\hbar^2}\frac{1}{\beta}\int_{-\beta\mu}^{\infty}\frac{dz}{e^{z}+1}=
S\frac{m}{\pi\hbar^2\beta}\left(\int_{-\beta\mu}^{0}\frac{dz}{e^{z}+1}+\int_{0}^{\infty}\frac{dz}{e^{z}+1}\right).
\label{N2}
\end{equation}
Using the identity
\begin{equation}
\frac{1}{e^{z}+1}=1-\frac{1}{e^{-z}+1}
\label{identity}
\end{equation}
in the first integral of Eq.~(\ref{N2}), one further obtains the following for the parenthesized expression above
\begin{equation}
\int_{-\beta\mu}^{0}\!\!\!dz-\!\int_{-\beta\mu}^{0}\frac{dz}{e^{-z}+1}+\!\int_{0}^{\infty}\!\!\!\frac{dz}{e^{z}+1}=
\beta\mu-\!\int_{0}^{\beta\mu}\!\!\!\frac{dz}{e^{z}+1}+\!\int_{0}^{\infty}\!\!\!\frac{dz}{e^{z}+1}\approx
\beta\mu-\!\int_{0}^{\infty}\!\!\!\frac{dz}{e^{z}+1}+\!\int_{0}^{\infty}\!\!\!\frac{dz}{e^{z}+1}=\beta\mu\,.
\label{N3}
\end{equation}
Here, the replacement $\beta\mu\!\rightarrow\!\infty$ in the last step is an approximation consistent with $T$ being less than $T_F\!\sim\!10^5\;$K, which amounts to neglecting exponentially small terms in asymptotic series expansions of Eqs.~(\ref{meanK}), (\ref{meanN}) and such~\cite{LandauStatPhys}.

Plugging Eq.~(\ref{N3}) into Eq.~(\ref{N2}) leads to
\begin{equation}
\langle N\rangle=S\frac{m\mu}{\pi\hbar^2}\,.
\label{Nfinal}
\end{equation}
This is the final result for the mean number of particles (electrons) $\langle N\rangle$ in the system defined initially by Eq.~(\ref{meanN}), yielding also
\begin{equation}
n=\frac{m\mu}{\pi\hbar^2},
\label{nfinal}
\end{equation}
as per Eq.~(\ref{n}), and
\begin{equation}
\mu=\frac{\pi n\hbar^2}{m},
\label{mu}
\end{equation}
accordingly.

To calculate the mean kinetic energy per particle $\langle K\rangle$, one starts from Eq.~(\ref{meanK}) to obtain after plugging Eq.~(\ref{ek}) in it
\begin{equation}
\langle K\rangle=\frac{1}{\pi n}\frac{m}{\hbar^2}\!\int_{0}^{\infty}\!\!\!\frac{d\epsilon\epsilon}{e^{\beta(\epsilon-\mu)}+1},
\label{K1}
\end{equation}
which after the substitution (\ref{substitution}) takes the form
\begin{equation}
\langle K\rangle=\frac{1}{\pi n}\frac{m}{\hbar^2}\frac{1}{\beta}\!\int_{-\beta\mu}^{\infty}dz\,\frac{\mu+z/\beta}{e^{z}+1}=
\frac{m}{\pi n\hbar^2\beta}\left(\int_{-\beta\mu}^{0}dz\,\frac{\mu+z/\beta}{e^{z}+1}+\int_{0}^{\infty}\!\!\!dz\,\frac{\mu+z/\beta}{e^{z}+1}\right).
\label{K2}
\end{equation}
Here, as before, the parenthesized expression can be transformed using Eq.~(\ref{identity}) to obtain
\begin{eqnarray}
\int_{-\beta\mu}^{0}\!\!\!dz\Big(\mu+\frac{z}{\beta}\Big)
-\!\int_{-\beta\mu}^{0}\!\!\!dz\,\frac{\mu+z/\beta}{e^{-z}+1}+\!\int_{0}^{\infty}\!\!\!dz\,\frac{\mu+z/\beta}{e^{z}+1}=
\int_{0}^{\beta\mu}\!\!\!dz\Big(\mu-\frac{z}{\beta}\Big)
-\!\int_{0}^{\beta\mu}\!\!\!dz\,\frac{\mu-z/\beta}{e^{z}+1}+\!\int_{0}^{\infty}\!\!\!dz\,\frac{\mu+z/\beta}{e^{z}+1}\hskip0.5cm\nonumber\\
\approx\frac{\beta\mu^2}{2}-
\!\int_{0}^{\infty}\!\!\!dz\,\frac{\mu-z/\beta}{e^{z}+1}+\!\int_{0}^{\infty}\!\!\!dz\,\frac{\mu+z/\beta}{e^{z}+1}=
\frac{2}{\beta}\Big[\Big(\frac{\beta\mu}{2}\Big)^{\!2}\!+\!\int_{0}^{\infty}\!\!\!dz\,\frac{z}{e^{z}+1}\Big]=
\frac{2}{\beta}\Big[\Big(\frac{\beta\mu}{2}\Big)^{\!2}\!+\!\frac{\pi^2}{12}\Big],\hskip1.5cm
\label{K3}
\end{eqnarray}
where to calculate the remaining integral in the last step, the following analytical formula in terms of Bernoulli numbers is used~\cite{LandauStatPhys}
\[
\int_0^\infty\!\!\!\!\!dz\frac{z^{2n-1}}{e^z+1}=\frac{2^{2n-1}-1}{2n}\pi^{2n}B_n,\;\;\;B_1=\frac{1}{6},\;B_2=\frac{1}{30},\;B_3=\frac{1}{42},\;B_4=\frac{1}{30},\;\cdots
\]

Plugging Eq.~(\ref{K3}) into Eq.~(\ref{K2}) and eliminating $\mu$ by using Eq.~(\ref{mu}), one obtains the following final form of Eq.~(\ref{meanK}) defining the mean kinetic energy per particle in the system
\begin{equation}
\langle K\rangle=\frac{\pi n\hbar^2}{2m}\!\left[1\!+\!\frac{1}{3}\Big(\frac{m}{n\hbar^2\beta}\Big)^{\!2}\right].
\label{Kfinal}
\end{equation}

\section{Mean kinetic energy per particle at high temperatures}

The high-temperature regime is used in the main text with the only purpose to introduce the $T_c$ scaling unit. This is the classical regime, for which the classical energy equipartition theorem says that the mean kinetic energy per particle $\langle K\rangle$ in a many-particle system of weakly interacting particles is equal to $k_BT/2$ multiplied by the number of degrees of freedom of the particles in the system, thus yielding  $\langle K\rangle\!=\!k_BT$ in our case. Below, for completeness of our analysis, this is demonstrated by direct analytical calculations.

In the high-$T$ regime, the Fermi-particle distribution function (\ref{nF}) becomes
\begin{equation}
n_F(\textbf{k})=\frac{1}{e^{\beta[\epsilon(\textbf{k})-\mu]}+1}\approx e^{\beta\mu}e^{-\beta\epsilon(\textbf{k})},
\label{nFhighT}
\end{equation}
and Eq.~(\ref{meanK}) in view of Eqs.~(\ref{meanN}) and (\ref{ansatz}) takes the form
\begin{equation}
\langle K\rangle=\frac{\sum_{\textbf{k}}\epsilon(\textbf{k})e^{-\beta\epsilon(\textbf{k})}}{\sum_{\textbf{k}}e^{-\beta\epsilon(\textbf{k})}}=-\frac{\partial}{\partial\beta}\ln\sum_{\textbf{k}}e^{-\beta\epsilon(\textbf{k})}
=-\frac{\partial}{\partial\beta}\ln\frac{Sm}{2\pi\hbar^2}\!\int_0^\infty\!\!\!d\epsilon\,e^{-\beta\epsilon}=-\frac{\partial}{\partial\beta}\ln\frac{Sm}{2\pi\hbar^2\beta}=\frac{1}{\beta}
\label{meaKhig}
\end{equation}
as expected.

\section{Domain of definition of the melting surface equation}

The melting surface equation presented in Eq.~(7) of the main text has the form
\begin{equation}
t=\frac{2}{\pi}\sqrt{3\nu\big[F(d,\varepsilon,\varepsilon_{1,2})-\nu\big]},
\label{MeltingSurfaceEq}
\end{equation}
where $\nu\!=\!n/n_c$, $t\!=\!T/T_c$, and $F(d,\varepsilon,\varepsilon_{1,2})$ is the dimensionless function of actual TD film parameters. Its projection on the $d=0$-plane of the $(d,\nu,t)$ three-coordinate space takes the form
\begin{equation}
t=\frac{2}{\pi}\sqrt{3\nu\Big(\frac{2\sqrt{\nu}}{\varepsilon_1\!+\varepsilon_2}-\nu\Big)},
\label{zero-d}
\end{equation}
which turns into the PF melting curve
\begin{equation}
t=\frac{2}{\pi}\sqrt{3\nu\big(\!\sqrt{\nu}-\nu\big)}
\label{zero-d-eps1}
\end{equation}
when $\varepsilon_1\!=\varepsilon_2\!=\!1$.

\begin{figure}[p]
\hskip-0.35cm\includegraphics[width=0.55\textwidth]{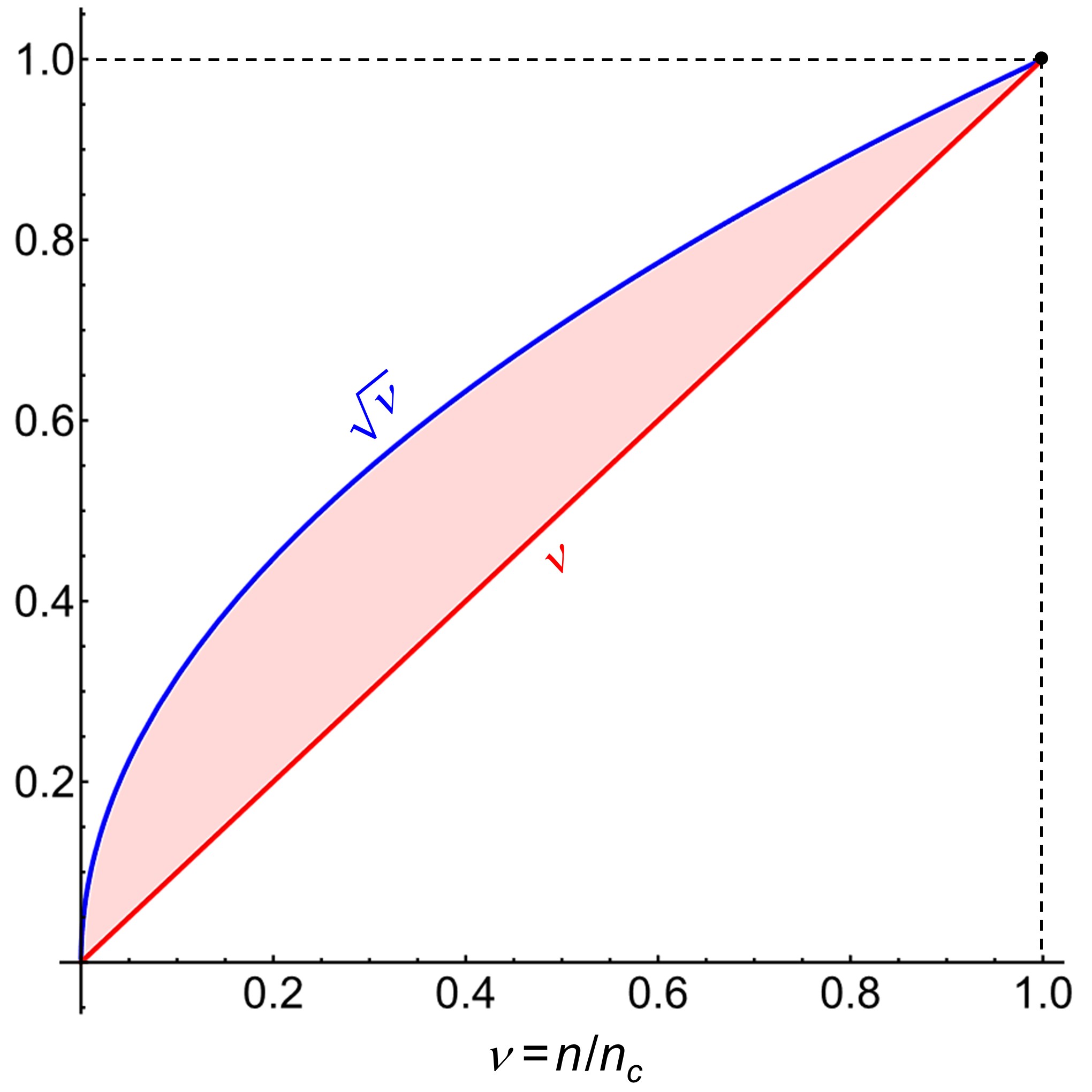}
\caption{The domain of the PF model melting curve (\ref{zero-d-eps1}) is the light-red shaded area bounded by the functions $\sqrt{\nu}$ and $\nu$ from the top and bottom, respectively.}
\label{fig4s}
\end{figure}

\begin{figure}[p]
\hskip-0.35cm\includegraphics[width=0.7\textwidth]{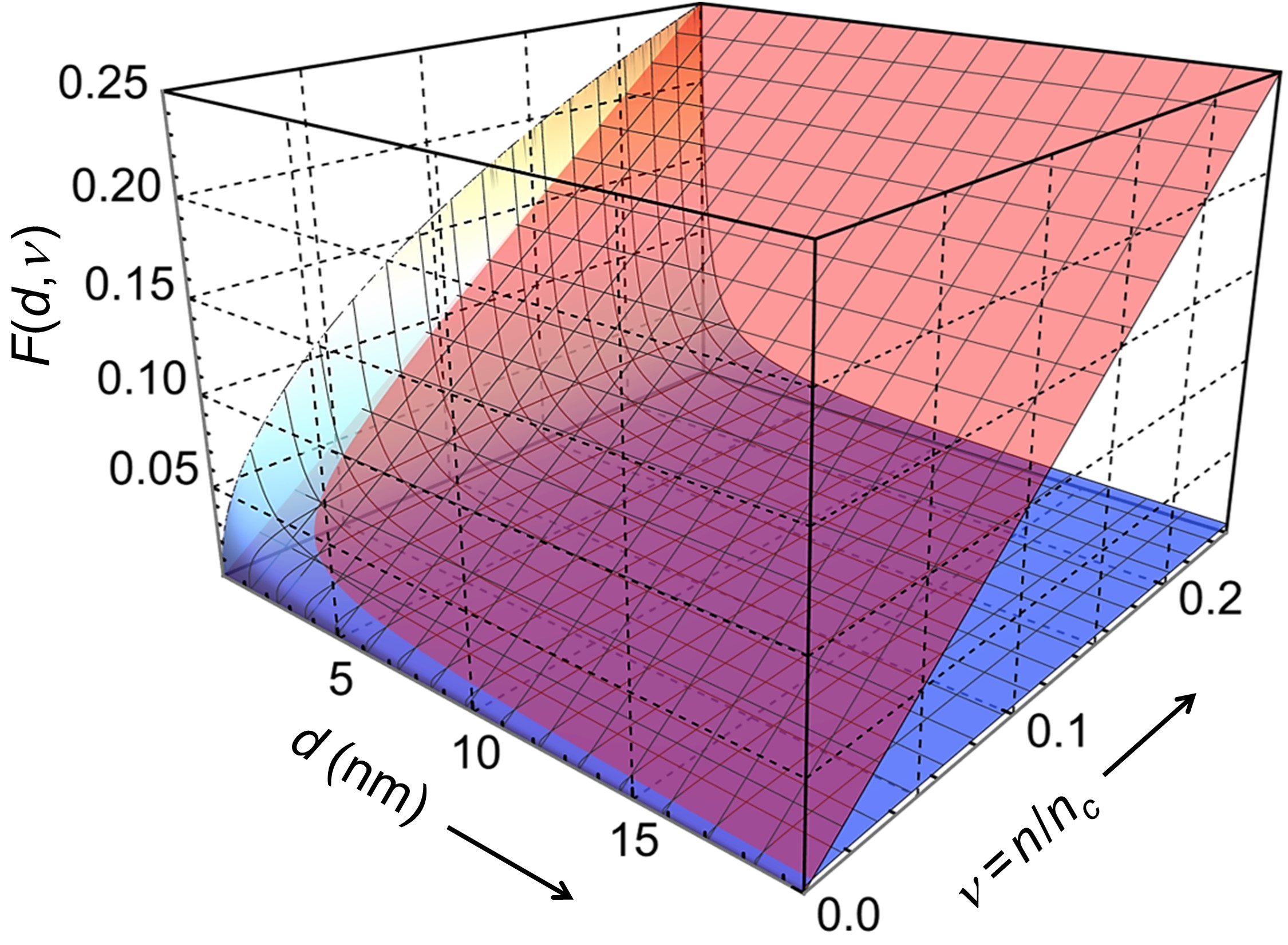}
\caption{The domain of the general melting surface equation (\ref{MeltingSurfaceEq}) lies above the $F\!=\!\nu$-plane (light-red) in the $(d,\nu,F)$ three-coordinate space. Shown here is the graph calculated for the air($\varepsilon_1\!=\!1$)/TiN($\varepsilon\!=\!9$)/MgO($\varepsilon_2\!=\!3$) TD system.}
\label{fig5s}
\end{figure}

Figure~\ref{fig4s} shows the domain of Eq.~(\ref{zero-d-eps1}). This is the light-red shaded area bounded by the functions $\sqrt{\nu}$ and $\nu$ from the top and bottom, respectively. Figure~\ref{fig5s}, calculated for the air($\varepsilon_1\!=\!1$)/TiN($\varepsilon\!=\!9$)/MgO($\varepsilon_2\!=\!3$) TD system as an example, shows the domain of Eq.~(\ref{MeltingSurfaceEq}). This can be seen to be the segment of 3D space above the $F\!=\!\nu$-plane. It can be seen that for quite a broad range of $d$, or screening lengths $r_0\!=\!\varepsilon d/(\varepsilon_1\!+\varepsilon_2)$, there is always a solution to guarantee the Wigner solid phase in the vicinity of $\nu\!\gtrsim\!0$, while in the vicinity of $\nu\!\lesssim\!0.25$, or more generally $\nu\!\lesssim\!4/(\varepsilon_1\!+\varepsilon_2)^2$ representing the right boundary of the domain of Eq.~(\ref{zero-d}), solutions are only possible for $d$ low enough. These solutions are strongly dependent on the permittivities of substrate and superstrate materials and to a lesser extent on $d$ itself. As a consequence, by choosing an appropriate substrate material it is possible to shift the right domain boundary of Eq.~(\ref{zero-d}) closer to the right domain boundary $\nu\!=\!1$ of the idealized PF model shown in Fig.~\ref{fig4s}. For example, by choosing a substrate with $\varepsilon_2\!=\!2$ (teflon) instead of 3, one shifts the TiN right domain boundary to $\nu\!=\!(2/3)^2\!\approx\!0.44$. This pushes the domain of Eq.~(\ref{zero-d}) out to the right by almost twice to give $n\!=\!0.44n_c\!\approx\!2.2\!\times\!10^{15}\,$cm$^{-2}$ for the electron density at the right boundary of the Wigner solid phase, in which case the ultrathin TiN system with $d\!\lesssim\!1\;$nm enters (or is about to enter) the solid phase region from the right in the very same way the $2\;$nm-thick air/HfN/MgO TD system does as shown in the main text.

Surface electron density lowering and simultaneous shifting of the Wigner solid phase boundary towards higher electron densities due to the thickness reduction and proper choice of substrate materials, open up the opportunity to cross into the Wigner solid phase region through the higher electron density boundary $\nu\!\lesssim\!4/(\varepsilon_1\!+\!\varepsilon_2)^2$. This makes TD systems advantageous in studies of strong electron correlation phenomena such as Wigner crystallization at elevated temperatures. This is an advantage over 2D systems such as monolayer and quasi-monolayer semiconductor TMDC materials~\cite{Smolenski,Park}, which are typically restricted to work at electron densities near the left boundary $\nu\!\gtrsim\!0$ of the Wigner solid phase region and so at cryogenic temperatures as explained in the main text.

\section{Temperature-dependent PF ratio of TD films}

The generalized PF ratio presented in Eq.~(5) of the main text can be rewritten as follows
\begin{equation}
\Gamma(d,\nu,t,\varepsilon,\varepsilon_{1,2})=\Gamma_0\,\frac{12\nu F(d,\varepsilon,\varepsilon_{1,2})}{12\nu^2+\pi^2t^2},
\label{PFratioGen}
\end{equation}
which in the limit of $d\!\rightarrow\!0$ takes the form
\begin{equation}
\Gamma(\nu,t,\varepsilon,\varepsilon_{1,2})=\Gamma_0\,\frac{24\nu^{3/2}}{(\epsilon_1+\epsilon_2)(12\nu^2+\pi^2t^2)}
\label{PFratioEps}
\end{equation}
to give
\[
\Gamma(\nu,t)=\Gamma_0\,\frac{12\nu^{3/2}}{12\nu^2+\pi^2t^2}
\]
for the idealized 2D electron gas system free-standing in air. Notably, it can be seen from these equations that while converging to zero for $\nu\!\rightarrow\!0$ and $\nu\!\rightarrow\!+\infty$ at all non-zero $T$ and at all $T$, respectively, with maximum in between at $\nu_m\!=\!\pi t/2$ raising up as $T$ decreases, they all are divergent as $1/\!\sqrt{\nu}$ for $\nu\!\rightarrow\!0$ in the artificial case of the absolute zero of temperature. While the dependence of the limiting value on the path taken is not surprising for a function of two variables, in reality the electrostatic repulsion tends to zero at very low $\nu$ and kinetic energy per particle remains finite due to quantum fluctuations. Therefore, the potential-to-kinetic energy ratio must go to zero for $\nu\!\rightarrow\!0$ at all $T$, including $T\!=\!0\,$K. This is in sharp contrast to zero-$T$ theory predicting more favorable crystallization conditions for $\nu\!\rightarrow\!0$~\cite{Tanatar89}, which is why these predictions should not be taken for granted.

\begin{figure}[p]
\hskip-0.35cm\includegraphics[width=0.7\textwidth]{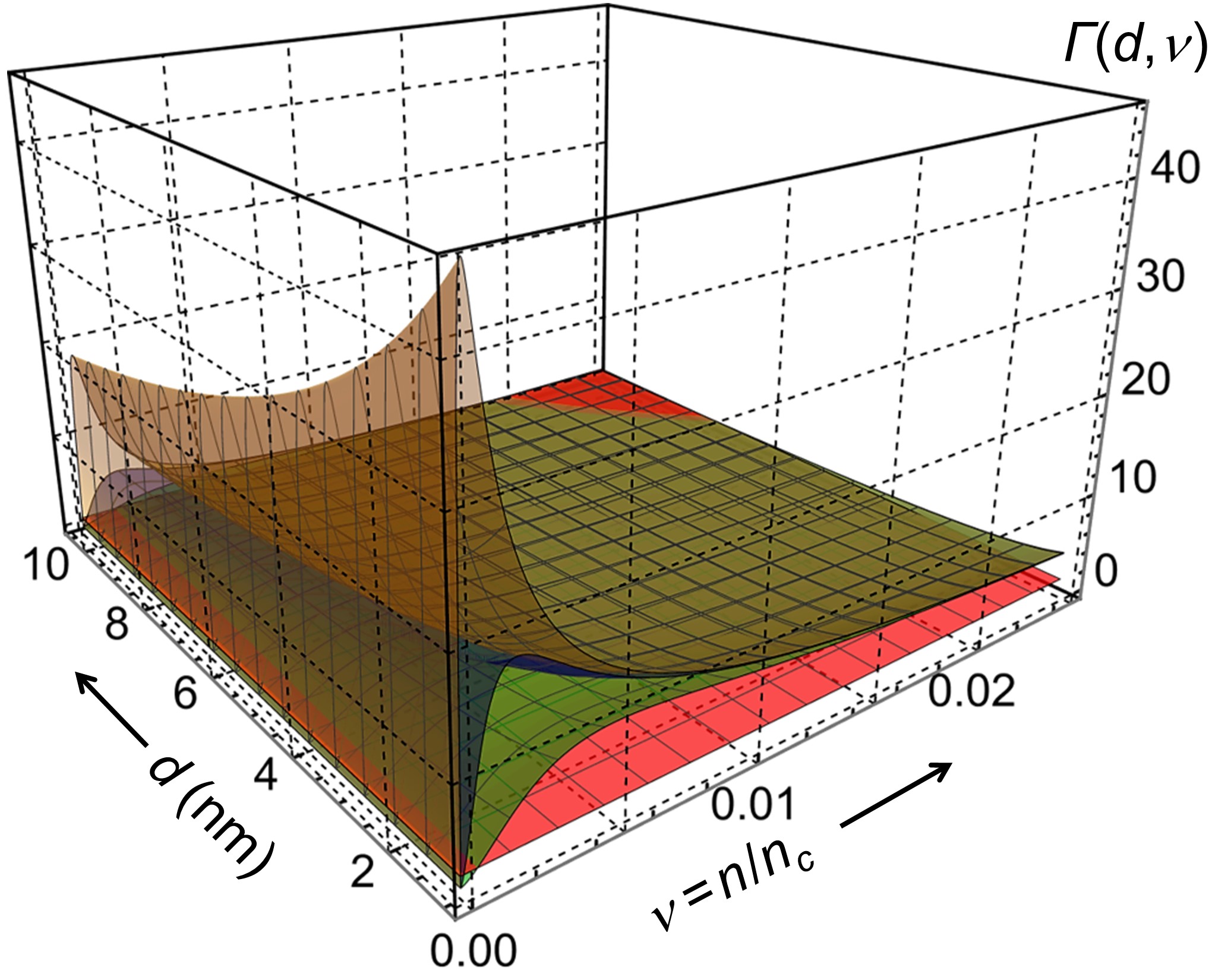}
\caption{The $\Gamma(d,\nu)$ surfaces calculated from Eq.~(\ref{PFratioGen}) for the air($\varepsilon_1\!=\!1$)/TiN($\varepsilon\!=\!9$)/MgO($\varepsilon_2\!=\!3$) TD system at temperatures $T\!=\!20\;$K (dark yellow), $100\;$K (blue), and $300\;$K (green) for $\nu$ in the domain corresponding to $n\!\sim\!10^{13}\!\div\!10^{14}\,$cm$^{-2}$. Red plane at the bottom is the $\Gamma\!=\!1$ plane.}
\label{fig6s}
\end{figure}

\begin{figure}[p]
\hskip-0.35cm\includegraphics[width=0.7\textwidth]{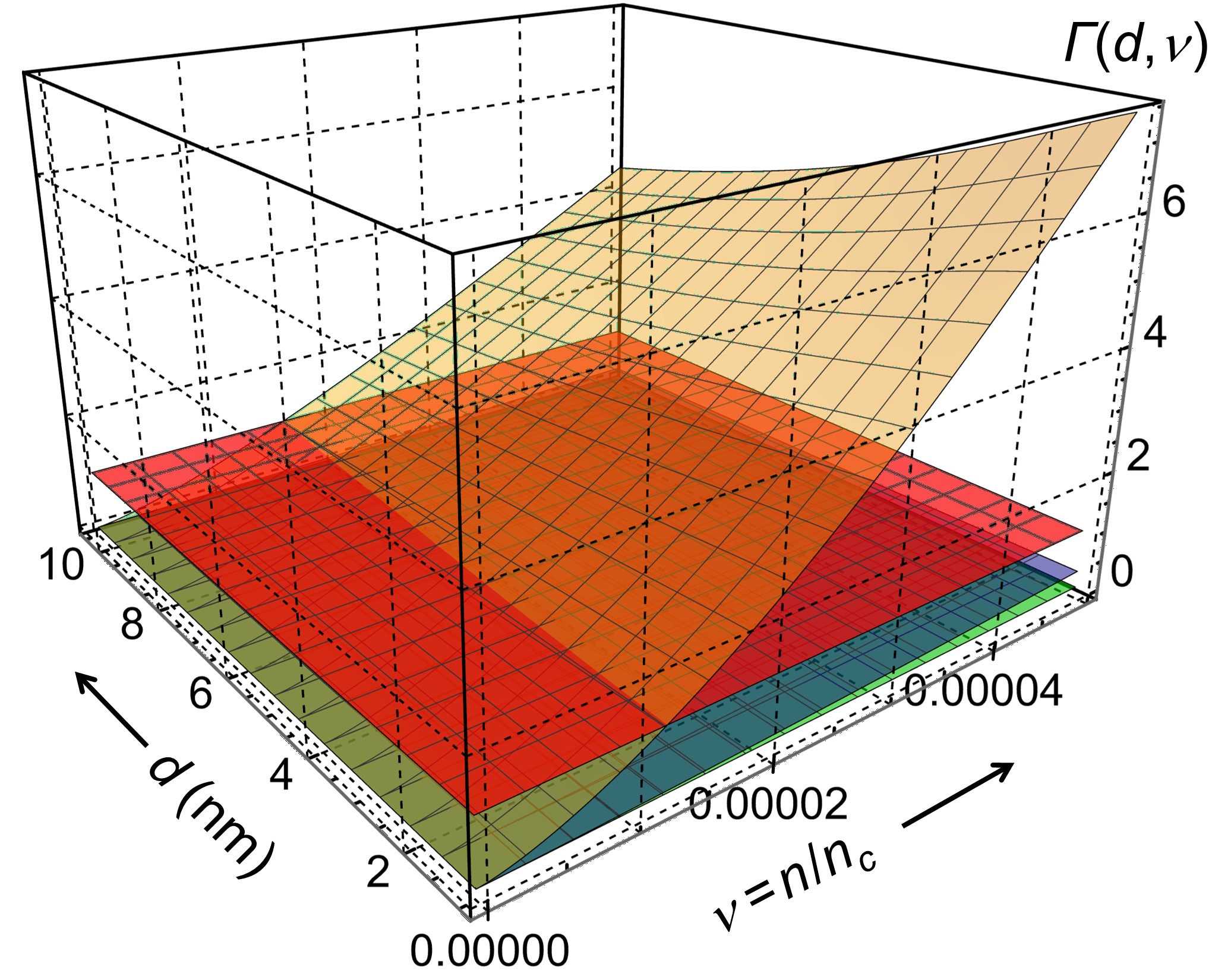}
\caption{Same as in Fig.~\ref{fig6s} plotted in the domain of extremely low $\nu$ corresponding to electron densities $n\!\sim\!10^{11}\,$cm$^{-2}$.}
\label{fig7s}
\end{figure}

Figure~\ref{fig6s} shows the $\Gamma(d,\nu)$ surfaces calculated from Eq.~(\ref{PFratioGen}) for the air($\varepsilon_1\!=\!1$)/TiN($\varepsilon\!=\!9$)/MgO($\varepsilon_2\!=\!3$) TD system at $T\!=\!20\;$K (dark yellow), $100\;$K (blue), and $300\;$K (green) for $\nu$ in the domain corresponding to $n\!\sim\!10^{13}\!\div\!10^{14}\,$cm$^{-2}$ discussed in the main text. Red plane at the bottom is the $\Gamma\!=\!1$ plane. Figure~\ref{fig7s} shows the same surfaces in the domain of extremely low $\nu$ corresponding to electron densities $n\!\lesssim\!10^{11}\,$cm$^{-2}$ typical of quasi-2D semiconductor materials such as TMDC and GaAs heterostructures~\cite{SmolenskiPark,Shayegan99GaAs}. All three surfaces in Fig.~\ref{fig6s} can be seen to fulfil inequality $\Gamma\!\gtrsim\!10$ for $\nu\!\sim\!0.01$ (or $n\!\sim\!5\!\times\!10^{13}\,$cm$^{-2}$) and $d\!\lesssim\!1\!\div\!3\;$nm. This is more than enough to favor the electron Wigner crystallization effect at room $T$ and below.

In Figure~\ref{fig7s}, on the contrary, only the yellow surface ($T\!=\!20\;$K) can be seen being above the $\Gamma\!=\!1$ plane and the other two are well below. Moreover, it can be seen that by decreasing $\nu$, or by increasing $T$, one makes it go below the $\Gamma\!=\!1$ plane, too, which would lead to melting of the Wigner solid already formed. This explains the main signatures of electron Wigner crystallization previously reported for zero-magnetic field experiments both with quasi-monolayer TMDC semiconductors~\cite{SmolenskiPark} and with $p$-doped GaAs/AlGaAs heterostructures~\cite{Shayegan99GaAs}.

Thus, it is generally a mistake to think that by reducing the carrier density one would provide better conditions for Wigner solid formation in quasi-2D systems. As a matter of fact, this is only the case for TD metallic and semimetallic compounds whose original electron density is relatively high. By lowering it due to thickness reduction one makes the electron system enter the Wigner solid phase region ($\Gamma\!>\!1$) from the high $\nu$ side as Fig.~\ref{fig6s} shows. The intrinsic electron density of quasi-2D semiconductors is a few orders of magnitude lower. They are situated at the low density side ($\nu\!\sim\!0$) of the Wigner solid phase region, where at all finite temperatures, no matter how low they are, the reduction of the carrier density generally drives the system out of the Wigner solid phase region. This can be seen in Figs.~\ref{fig6s} and \ref{fig7s} as well as in the melting surface graphs presented in the main text.

